\documentclass[onecolumn,preprintnumbers,amsmath,amssymb,a4paper]{revtex4}

\usepackage{graphicx}
\usepackage{bm}

\begin{document}


\title{The Price of Neutrino Superluminality continues to rise}

\author{Arthur Hebecker}
\email{A.Hebecker@ThPhys.Uni-Heidelberg.de}

\author{Alexander Knochel}%
\email{A.K.Knochel@ThPhys.Uni-Heidelberg.de}
 \affiliation{Institut f\"ur Theoretische Physik, Philosophenweg 19, D-69120
Heidelberg, Germany}

\begin{abstract}
We revisit the model building challenges that one faces when trying to
reconcile the OPERA claim of neutrino superluminality with other
observational constraints. The severity of the supernova bound and of the
kinematical constraints of Cohen-Glashow type lead us to focus on
scenarios where all types of particles are superluminal inside matter.
In contrast to the Dvali-Vikman proposal, this matter effect needs to be very
short-ranged to avoid constraints from experiments on the Earth's
surface in low-density environments. Due to this short range, the
interaction underlying such a matter effect would have to be far stronger
than permitted by fifth-force bounds. As a conceivable way out we suggest to make
the matter effect ``binary'', i.e., dense matter does not directly
trigger superluminality, but merely induces the transition to a different phase
of some weakly coupled hidden sector. This phase exhibits spontaneous
Lorentz violation or at least a stronger than usual mediation of some residual
Lorentz violation to all matter. The effect has not been observed before since
we have never before been able to measure the velocity of high-energy particles
in dense matter with sufficient precision.
\end{abstract}

\maketitle

\noindent

\section{Introduction}
Earlier this year, the OPERA collaboration made the surprising claim
\cite{:2011zb} of an early arrival time of CNGS muon antineutrinos 
traversing 730 kilometers of rock on their way from CERN to Gran Sasso.
This corresponds to a neutrino superluminality of
\begin{equation}
\Delta\equiv v^2_\nu-1\sim 5\times 10^{-5}
\end{equation}
at $6\sigma$ significance. The CNGS neutrinos have an average energy of $17$ 
GeV with a broad distribution reaching up to several tens of GeV, and a 
separate measurement of neutrinos above and below $20$ GeV has revealed no 
significant energy-dependence of the superluminality in this energy range. This
result was preceded by a weaker claim from MINOS \cite{Adamson:2007zzb}, and 
was recently confirmed in a followup investigation by OPERA using short 
bunches. The latter was carried out in order to exclude potential problems 
which might have arisen because the extractions used in the original 
measurement were much longer ($\sim 3000$ meters) than the observed effect 
($\sim 18$ meters). For the purpose of this note, we take the experimental 
result at face value and confront the model building challenges which it 
entails. In other words, we ignore any possible experimental problems 
and focus on the intellectual challenge of devising at least an effective 
field-theoretic framework in which the effect can be consistently described. 
Note that the corresponding effect in standard general relativity which arises 
due to the presence of the earth, is smaller than
the observed effect by a factor of $10^{-5}$ and thus negligible 
\cite{Kehagias:2011cb,Lust:2011fx,Alles:2011wq}.

Since the publication of \cite{:2011zb}, a large number of preprints has
appeared on the arXiv addressing or attacking aspects of the claim
or proposing models. The latter can roughly be categorized into models
of explicit Lorentz violation \cite{Alexandre:2011bu}, geometric solutions 
in extra dimensions \cite{Gubser:2011mp}, spontaneous Lorentz breaking 
\cite{Nojiri:2011ju}, deformed special relativity 
\cite{AmelinoCamelia:2011bz}, environmental superluminality \cite{
Dvali:2011mn,Kehagias:2011cb,Oda:2011kh}, and combinations of these ideas.

The first obvious phenomenological challenge is the observation of neutrinos 
from the supernova SN1987A a few hours before the optical confirmation 
\cite{Alexandre:2011bu}. The naive assumption of a constant and 
energy-independent superluminality $\Delta$ would have had these neutrinos 
reach Earth years before the photons. There are several ways out of this 
problem: 

\begin{itemize}

\item The superluminality is sufficiently energy dependent such that the 
low-energy neutrinos from SN1987A see very little of it \cite{Ellis:2011uk,
Dass:2011yj,Li:2011ue,AmelinoCamelia:2011dx,Cacciapaglia:2011ax}.

\item It is flavor dependent and electron neutrinos are not superluminal, 
while muon neutrinos are. This appears to seriously interfere neutrino 
oscillations \cite{Fargion:2011hd}.

\item The velocity of neutrinos is location dependent, e.g. a function of
the matter density. This way, neutrinos are barely superluminal in
interstellar space, whereas the CNGS neutrinos travelling close to or inside
Earth, are \cite{Dvali:2011mn,Kehagias:2011cb,Oda:2011kh}.
\end{itemize}

The to date most severe direct constraints come from the modified
kinematics of decays involving superluminal particles, in particular the
argument by Cohen and Glashow (CG) that superluminal neutrinos can radiate
non-superluminal particles, such as $\nu\rightarrow e^+ e^- \nu$, and thus 
quickly lose energy \cite{Cohen:2011hx,Bi:2011nd,Li:2011ad}. The energy threshold in 
terms of the superluminality is given by 
\begin{equation}
E_\nu > \frac{\sum m_i}{\sqrt{\Delta}}\,,\label{cgthreshold} 
\end{equation}
where $m_i$ are the masses of the final state particles. Likewise, high-energy 
mesons would not have any phase space left to decay when neutrinos are in the 
final state, thus eliminating both the production of atmospheric neutrinos 
\cite{GonzalezMestres:2011jc,Cowsik:2011wv} and of the CNGS neutrino beam 
itself.
One can try to evade this problem as follows:
\begin{itemize}
\item Lorentz symmetry is not broken but deformed, alternative momentum and 
energy conservation relations hold, and the CG effect is avoided
\cite{AmelinoCamelia:2011bz,Klinkhamer:2011js}. Our only excuse for not 
following this line of thinking at the moment is our insufficient understanding 
of the underlying field-theoretic framework.
\item Superluminality is achieved through a nontrivial dispersion relation which
manages to suppress the CG effect, see for example \cite{Mohanty:2011rm}.
However, we need to suppress the CG effect also for energies $E\gg 17$ GeV, 
e.g. to allow for high-energy neutrinos traversing the Earth, as they are 
observed in ``upward'' events by IceCube \cite{icecube} (see also
\cite{Bi:2011nd}). This appears to be 
difficult. \item All particle species are equally superluminal, which amounts to a 
rescaling of the energy and leads to an effectively Lorentz-invariant
kinematics with this rescaled energy. Note that (\ref{cgthreshold}) implies 
that superluminal neutrinos above $\sim 40$ GeV have enough energy to
produce pion pairs via neutral currents, $\nu\rightarrow \nu \pi
\pi$, making mere electron superluminality insufficient to avoid the CG effect.
 \end{itemize}

In the following we retain standard energy-momentum conservation in field 
theory. Thus, from the considerations above, we are compelled to consider the 
following picture: {\it Neutrinos are superluminal only when close to dense 
matter. The effect originates from a modified dispersion relation 
which they share with all other particle species.} The severe constraints, 
e.g. on electron superluminality in synchrotrons \cite{Altschul:2006ka,
Altschul:2009xh}, demand that this effect has a very short range, i.e. that
superluminality goes away millimeters or less outside of solids.

\section{Matter-Dependent Superluminality}
An elegant way to produce neutrino superluminality close to the Earth 
in a completely Lorentz-invariant setting was proposed by Dvali and Vikman
\cite{Dvali:2011mn}. DV exploit the effect that the mere presence of the Earth
constitutes an effective violation of Lorentz invariance, which is then
communicated to the neutrinos via a tensor fifth force. The inverse mass of this
tensor $h_{\mu\nu}$ is dialed to a value between the radius of the Earth and 
the size of the solar system such that it is effectively massless on Earth. 
The proposed couplings of the tensor to other fields are via the 
energy-momentum tensor, but are nonuniversal between neutrinos and other 
particles:
\begin{equation}
\mathcal{L}\supset \frac{h_{\mu\nu}}{M_*} \overline\nu i
\partial^{\mu}\gamma^{\nu}\nu + \frac{h_{\mu\nu}}{M} T^{\mu\nu}_{\nu\!\!\!/}.
\end{equation}
The net superluminality on the Earth's surface is given by the dispersion 
relation $p_{\mu}p^{\mu}=\vec{p}^2\, \Delta$, where \cite{Dvali:2011mn}
\begin{equation}
v_\nu-c=\Delta/2=-\frac{M_E}{4 \pi M_* M R_E}\,.
\end{equation}
Clearly, we need $M_* M <0$ in order to produce {\it super}luminality.
This model, as it stands, avoids the constraints from SN1987A, but not the CG
effect. Also, it requires considerable fine tuning of counterterms in
order to cancel the 1-loop contributions e.g. to electron superluminality
\cite{Giudice:2011mm}. One could consider making the tensor heavy at the 
inverse micron scale
in order to allow universal superluminality (which relaxes the fine tuning and
avoids CG at the same time), but apart from requiring a rather 
low scale $M_* M$, the relative sign between the couplings seems to make this
impossible - there is no obvious way to modify hadron dispersion relations in 
the direction of superluminality while hadrons at the same time provide the 
source of the tensor field. One way out might be to couple to $B-L$ or $B$, but
we are not aware of a viable realization of this idea using tensors. 
This problem seems to indicate that we need a further ingredient in order to
be compatible with SN1987A and the CG effect. We aim to disentangle the source
of Lorentz violation and the origin of matter dependence. Somewhat related 
ideas have appeared, partially in the form of comments, in
\cite{Evslin:2011vq,Ciuffoli:2011ji}.

\section{Effective Models}
To make our argument as general as possible, we introduce a spurion-like
Lorentz-violating dimensionless symmetric tensor which we choose diagonal in the
Earth frame for simplicity,
$\theta_{\mu\nu}=\mbox{diag}(\alpha,\beta,\beta,\beta)$. This tensor could be sourced
by the Earth, could represent a cosmological background, or could be of yet
another unknown origin. The scalar field
$\phi$ is sourced by matter and thus produces the matter dependence.
For the sake of concreteness, we consider the simple effective
Lagrangian
\begin{equation}
\mathcal{L}\supset\frac{\phi}{M} \,\theta_{\mu\nu} T^{\mu\nu} +\frac{1}{2}
\partial_\mu\phi \partial^\mu\phi- \frac{m^2}{2}\phi^2
\end{equation}
where $T^{\mu\nu}$ is now the complete energy-momentum tensor. In the spirit of
\cite{Dvali:2011mn},  $\langle\phi\rangle \theta_{\mu\nu}/M$ can be viewed as a
perturbation away from Minkowski of the effective
metric in which particles inside matter propagate, thus modifying their
dispersion relations. At the same time, this term sources the field $\phi$. We
choose $m^{-1}<10^{-3}\mbox{ meters}$, which implies that inside Earth (more
than a distance of $m^{-1}$ away from the surface),  we have
\begin{equation}\langle\phi\rangle_{\scriptsize E}=\frac{\theta_{\mu\nu}\langle
T^{\mu\nu}\rangle_{\mbox{\scriptsize E}} }{M  m^2}=\frac{\alpha \rho_E}{M
m^2}.
\end{equation}
We can now reinsert this into $\mathcal{L}$ in order to obtain an effective
Lagrangian for high energy particles inside matter,
\begin{equation}
\mathcal{L}_{eff}\supset \epsilon\,\left(\alpha^2\, T^{00} - \alpha \beta\, 
\eta_{ij} T^{ij} \right),\qquad \epsilon \equiv \frac{\rho_E}{M^2
m^2}\,.
\label{effective1}
\end{equation}
There are several ways to derive $\Delta$ from (\ref{effective1}). One can 
treat it as a contribution to the effective metric, or one can use it as an 
operator insertion into the decay processes. For neutrinos for example, the 
corresponding Feynman rule reads $i \epsilon (\alpha^2 \gamma^0 p^0 - \alpha
\beta \eta_{ij}\gamma^i p^i)$.  When Dyson-resummed for the external legs, 
this yields the modified, superluminal propagator. Both yield the same 
result, namely (for a massless particle) 
\begin{equation}
(1+\epsilon \alpha^2) p_0^2 - (1-\epsilon \alpha\beta) \vec{p}^2=0
\end{equation}
and thus $\Delta\approx -\epsilon (\alpha \beta +\alpha^2)$. What we have 
gained is the freedom to choose $\theta$ such that $\alpha\beta+\alpha^2=-1$, 
and thus $\Delta=\epsilon$. Note that this was impossible in the DV model 
with universal couplings since there, the corresponding expression would be 
a square and hence positive. We now get a relation between the range $m^{-1}$ 
and the coupling scale $M^{-1}$,
\begin{equation}
M=\sqrt{\frac{\rho_E}{\Delta}}\, m^{-1} \sim 3\cdot 10^9 \mbox{ GeV} \times
m^{-1}/\mbox{meter}.
\end{equation}
Thus, we obtain a Yukawa-type short-range modification of gravity with a 
suppression scale $M\sim 3\times 10^3$~GeV for $m^{-1}\sim \mu m$, in gross
conflict with precision experiments such as those by Lamoreaux {\it et al.} 
(\cite{Bordag:2001qi}, fig. 28). Extreme choices which one might consider in 
order to escape this are the nanometer range, but then $GeV$ suppressed 
interactions of a light scalar would have shown up in particle physics, or in 
the centimeter range, where however the suppression of the coupling is not 
strong enough to evade E\"otv\"os-type experiments. This model makes some 
assumptions, but the general problem of scales will be the same for any 
source of Lorentz violation, and will not be radically different for DV 
type models involving vector bosons rather than a tensor. Modifications of 
the scalar model by $D>5$ operators do not seem promising. There is the 
possibility to add a term of the form $\phi^2 T$ in order to reduce the 
effective mass of the scalar inside matter, but this must be finely tuned, 
and it would lead to a runaway potential in more dense matter. As yet another 
option, one might choose $|\alpha|\ll 1$ and $|\beta|\gg 1$, but the 
required $\beta$-values appear to be too extreme.

\section{Two phase models}
We want to retain the idea of a matter effect, but gain more freedom concerning
the coupling strength of the underlying scalar field. This might be doable in a
two-phase model, where the matter and the vacuum phase have roughly independent 
Lagrangian parameters. As a very simple attempt at such a model using
renormalizable couplings, consider the scalar potential
\begin{equation}
V(\phi)=\frac{\lambda}{4}\left[\left(\phi-\frac{\mu}{\sqrt{\lambda}}\right)^2-\frac{\mu^2}{\lambda}\right]^2
+ \phi^2 \mu^2 \delta
\label{mexicanhat}
\end{equation}
which, for $0<\delta \ll 1$, is simply a deformed double well with the global minimum
$V(0)=0$
and a local one $V(\Delta \phi)=4\delta \mu^4/\lambda$ at $\Delta\phi\equiv 2 \mu/\sqrt{\lambda}$. 
We assume that the presence of matter density tilts this potential, making the
``superluminal'' mininum at $\Delta \phi$ the global one. In order to achieve this, we postulate
the Lagrangian
\begin{equation}
\mathcal{L}\supset -V(\phi)-\frac{\phi}{\Lambda_{LV}} \theta_{\mu\nu}T^{\mu\nu}
+ \frac{\phi}{\Lambda_{LI}} T_\mu^\mu\label{twophaselagrangian}
\end{equation}
where we have a Lorentz invariant coupling producing the deformation of the
potential, and the Lorentz violating one inducing the superluminality.

Let $b$ denote the thickness of a domain wall separating the two phases.
We call the height of the corresponding potential barrier $V$.
The potential energy density contribution to the surface tension of the wall 
is approximately $\sigma \sim V b$, 
since the scalar field has to pass over the maximum inside the domain wall.
Likewise, the kinetic energy contribution is approximately $\sigma \sim b (\Delta
\phi/b)^2$. If we assume the optimal solution $Vb \sim \Delta \phi^2/b$, we
obtain an expression for the wall thickness, $b\sim\Delta\phi/\sqrt{V}$. Two
competing effects, the difference in energy density and the surface tension,
give us a critical bubble size inside or outside of matter via $dE =4\pi
\Delta V R^2 dR -8 \pi \sigma  R dR$. In our order of magnitude estimate, the
critical bubble size inside(outside) of matter is thus $R_c\sim \sigma/\Delta
V^{(\prime)}\sim b V/\Delta V^{(\prime)} \sim \Delta \phi
\sqrt{V}/\Delta V^{(\prime)}$. Here, $\Delta V$ and $\Delta V'$ denote the 
potential energy differences between the two local minima inside and outside
of matter respectively.
 
We can now make order of magnitude estimates for the parameters. 
The wall thickness $b$ in this model is 
simply given by $b\sim \Delta \phi/\sqrt{V} \sim \mu^{-1}$, and the height of the
potential well by $V \sim V(\Delta\phi/2)\sim \mu^4/\lambda$, and thus the
surface tension is $ \sigma \sim \mu^3/\lambda$. We demand that 
the range of the fifth force, and thus the wall thickness, are small enough 
to evade the experimental constraints from short range measurements
\cite{Bordag:2001qi}, and choose $\mu \sim (10^{-10}\mbox{ meter})^{-1}\sim
10^{-5} \mbox{ GeV}$. We can see from (\ref{twophaselagrangian})
and the required amount of superluminality, that $\Delta \phi/\Lambda_{LV} \sim 5 \cdot 10^{-5}$ 
and thus $\Lambda_{LV}\sim 1\mbox{ GeV}/\sqrt{\lambda}$, telling us that $\lambda<10^{-6}$ to evade collider
bounds. The surface tension is thus $\sigma>10^6 \mu^3 \sim 10^{20}
\mbox{GeV}/\mbox{meter}^2$, which is a very large surface energy density even
in macroscopic terms. 
While we have not checked all variants of this model, we suspect that the
framework described above is too restrictive.

Hence, if we want to keep this idea of a phase transition, we need to 
allow for more general
potentials as well as more general couplings to matter. 
Instead of working with a simple renormalizable potential as before, 
we postulate a general potential with two local minima which is tilted 
by the presence of energy density such that the superluminal vacuum becomes the true
one. We study this scenario with a Lagrangian of the form
\begin{equation}
\mathcal{L}\supset -V(\phi)-\frac{\Delta V'}{\Delta \phi}\phi -f\left(\phi\right) T_\mu^\mu -
g\left({\phi}\right)
\theta_{\mu\nu}T^{\mu\nu}
\end{equation}
in which we have introduced a Lorentz invariant coupling and a Lorentz violating one
with coefficient functions $f$ and $g$ which we choose in order to make the model
phenomenologically viable. 
\begin{figure}
\begin{center}
\includegraphics[width=0.5\linewidth]{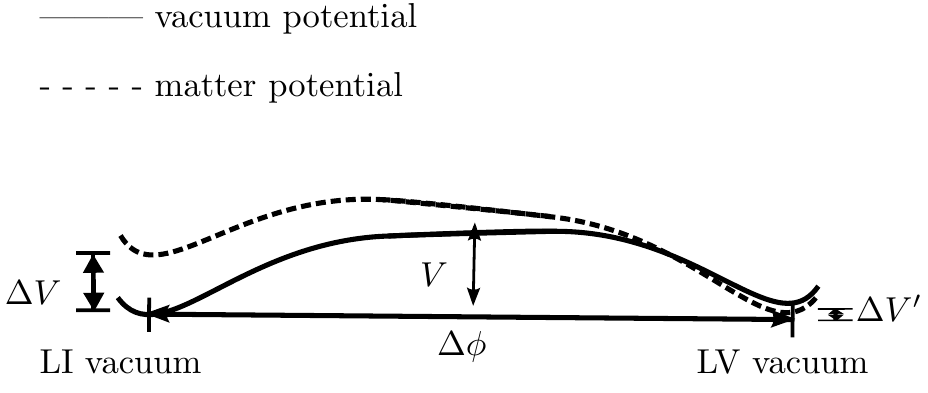}
\caption{A depiction of matter density dependent potentials. At low matter
density,
the Lorentz conserving local minimum is the true vacuum, while inside
matter, we have a superluminal vacuum. The energy density difference between the
true and false vacuum is given by $\Delta V'$ outside of matter, and by $\Delta
V$ inside.
\label{potential}}
\end{center}
\end{figure}
We assume a generic potential with $\Delta \phi$ the size of the
Vev given by the distance of the two local minima and $V$ the approximate potential barrier. 
For simplicity, we assume $V(0)=V(\Delta \phi)=0$, while
$\Delta V$ and $\Delta V'$, the energy differences between the
standard vacuum and the superluminal vacuum, arise from the choice of $f$.
This is illustrated in Figure \ref{potential}.

We now want to exploit our freedom of choosing $g$ and $f$ in order to make this
scenario compatible with OPERA and fifth force constraints. The perturbative 
couplings of $\phi$ to the Standard Model in the vacuum and in matter are given by $\partial^n f/\partial \phi^n(0),\, \partial^n
g/\partial \phi^n(0)$ and $\partial^n f/\partial \phi^n(\Delta \phi),\, \partial^n
g/\partial \phi^n(\Delta \phi)$ respectively. We thus want to set these to zero, 
e.g. by having $f$ and $g$ locally constant or of the type $e^{-1/(x-x_0)^2}$ at $x_0=0$ and
$x_0=\Delta
\phi/\Lambda$, with a smooth
transition between the two values. It is not obvious
whether or how such a model with a non-renormalizable non-analytic 
coupling can be UV completed in the hidden sector, and we postpone 
this discussion. For simplicity we choose $\theta_{00}=0$ in the Earth
frame such that
$g$ does not modify the potential while still inducing superluminality. 
In order to evade superluminality constraints
outside matter such as the supernova bound while reproducing the 
OPERA measurement, we need $g(0)=0$, $g(\Delta \phi) \sim 10^{-5}$. The function
$f$, which sources the deformation of the potential, must have a suitable
value $f(\Delta \phi)<0$ to produce the necessary shift $\Delta V\sim \rho_E \,f(\Delta
\phi) $ which makes $\phi=\Delta\phi$ the true vacuum inside matter. 
Further constraints on the function $f$ can be derived from the desire to have a
negligible contribution of the condensate to the mass of macroscopic objects,
i.e. the energy density of the true vacua inside and outside of matter should 
satisfy $|\Delta V' + \rho_E f(\Delta \phi)| \ll \rho_E$. The critical bubble of the
Lorentz conserving vacuum outside of matter should be smaller than e.g. the
beam pipe of synchrotrons, which are very sensitive to electron superluminality,
while the upper bound on the size of critical bubbles inside the Earth might be less 
strong. We furthermore require a thickness of the domain wall $b\gtrsim
10^{-9}\mbox{ meters}$, in order to have a homogeneous phase inside matter.

\section{Conclusions}

We have attempted to account for neutrino superluminality, as reported by OPERA,
while staying within the familiar framework of low-energy effective field
theory. Even if one is prepared to allow for explicit or spontaneous Lorentz
violation, this turns out to be surprisingly challenging: On the one hand, the
supernova bound strongly suggests that
we are dealing with a matter effect. On the other hand, the kinematical
constraints of Cohen-Glashow type force one to extend superluminality
to all types of particles. However, superluminality of various standard model
particles or a varying velocity of photons appears to be completely excluded by
a multitude of laboratory experiments. An obvious way out is to make
superluminality a universal, matter-induced effect, but with a very short
(sub-mm) range. One now faces the problem that only a small
volume of rock around each point of the neutrino trajectory contributes to the
effect. To compensate for this smallness, the short-ranged mediator
field has to couple rather strongly. This tends to be in conflict with
fifth-force bounds.

Given this situation, one is naturally led to consider a non-trivial
phase structure of some hidden sector: This hidden sector comes in a `matter
phase' associated with superluminality and a vacuum phase where
no such effect is present. Two couplings of the hidden sector to our
world are mandatory: A coupling to energy density, which stabilises
the superluminal phase inside matter, and a coupling to all Standard Model
kinetic terms, which induces superluminality whenever the
hidden sector is in that phase. The crucial point is that, in contrast to the
fifth-force-case above, these couplings do not need to be linear in any
dynamical scalar field. The constraints are now associated with
very different parameters: the energy density difference between the phases, the
domain-wall thickness, and its tension. As a further (derived) quantity, the
size of critical bubbles relevant in the transition between the phases may be
important in certain situations. As discussed in more detail in the main text,
there appears to be enough freedom to satisfy at least some of the most obvious
constraints. A more explicit construction and a more detailed analysis of
this suggested two-phase model is clearly necessary.
\section*{Acknowledgements}
We acknowledge helpful conversations with A. Vikman.
This work was partially supported by the Transregio TR33 ``The Dark Universe''.
\bibliography{superlumbib}
\end{document}